\documentclass{imsart}

\usepackage{orcidlink,thumbpdf,lmodern}
\usepackage{listings}
\usepackage[utf8]{inputenc}
\usepackage{natbib}
\usepackage{bm}
\usepackage{amsmath}
\usepackage{amssymb}
\usepackage{mathptmx}
\usepackage{tikz} 
\usetikzlibrary{shapes,arrows}

\newcommand{\appsim}{\stackrel{\mathrm{a}}{\sim}}

\usepackage{amsmath}

\begin{document}


\begin{frontmatter}
\title{{\bf csSampling}: An R Package for Bayesian Models for Complex
Survey Data}

\runtitle{Bayesian Models for Survey Data}

\begin{aug}
\author{\fnms{Ryan} \snm{Hornby}\thanksref{addr1}}
\and
\author{\fnms{Matthew R.} \snm{Williams}\thanksref{addr2}}
\and
\author{\fnms{Terrance D.} \snm{Savitsky}\thanksref{addr3}}
\and
\author{\fnms{Mahmoud} \snm{Elkasabi}\thanksref{addr4}}

\runauthor{Hornby, Williams, Savitsky \and Elkasabi}

\address[addr1]{Vassar College
    \href{mailto:ryanhornby1999@gmail.com}{ryanhornby1999@gmail.com}
}

\address[addr2]{RTI International
    \href{mailto:mrwilliams@rti.org}{mrwilliams@rti.org}
}
\address[addr3]{U.S. Bureau of Labor Statistics, Office of Survey Methods Research
    \href{mailto:Savitsky.Terrance@bls.gov}{Savitsky.Terrance@bls.gov}
}
\address[addr4]{RTI International
    \href{mailto:melkasabi@rti.org}{melkasabi@rti.org}
}

\end{aug}

\begin{abstract}
We present {\bf csSampling}, an R package for estimation of Bayesian models for data collected from complex survey samples. {\bf csSampling} combines functionality from the probabilistic programming language Stan (via the {\bf rstan} and {\bf brms} R packages) and the handling of complex survey data from the {\bf survey} R package. Under this approach, the user creates a survey-weighted model in {\bf brms} or provides a custom weighted model via {\bf rstan}. Survey design information is provided via the \texttt{svydesign} function of the {\bf survey} package. The \texttt{cs\_sampling} function of {\bf csSampling} estimates the weighted stan model and provides an asymptotic covariance correction for model mis-specification due to using survey sampling weights as plug-in values in the likelihood. This is often known as a ``design effect'' which is the ratio between the variance from a complex survey sample and a simple random sample of the same size. The resulting adjusted posterior draws can then be used for the usual Bayesian inference while also achieving frequentist properties of asymptotic consistency and correct uncertainty (e.g. coverage).
\end{abstract}


\end{frontmatter}

\hypertarget{introduction}{%
\section{Introduction}\label{introduction}}
Multipurpose statistical modelling packages such as {\bf brms} \citep{brms} provide a powerful suite of models and features for performing Bayesian analysis for multilevel models, while using syntax closely related to base R and popular random effects packages. The underlying Stan \citep{stan:2015} system can also be accessed through {\bf rstan} \citep{Rstan} and other packages, allowing for additional customization of models.

Our motivation is to extend the use of these models and their implementations to the case where data are collected under a complex survey sampling design associated with a finite population. For simple regression models (including proportional hazard models), the {\bf survey} package \citep{Rsurvey} allows for the incorporation of features of the survey sampling design (unequal probabilities of selection, stratification, and clustering) to provide pseudo-maximum likelihood estimation of regression model parameters for the underlying population. Our goal is to address more complicated, hierarchical statistical models for complex survey data that are not readily handled by the {\bf survey} package. 

This article describes {\bf csSampling}, an R package that implements the methods introduced in \cite{WilliamsISR21}. The {\bf csSampling} package implements a hybrid approach using Stan and the {\bf survey} package to produce consistent survey-weighted pseudo-posteriors with asymptotically correct uncertainty quantification, extending the capabilities of {\bf brms} and Stan to handle complex survey data.

\hypertarget{inference-from-complex-survey-samples}{%
\subsection{Inference from Complex Survey Samples}\label{inference-from-complex-survey-samples}}
We suppose that a data analyst seeks to estimate a Bayesian model, $P_{\theta_{0}}$, which is hypothesized as the generating model for observed data $Y$ for a population $U = (1,\ldots N)$ of size $N$. However, the analyst only has access to a selected sample $S = (1,\ldots, n \le N)$ drawn from the underlying population. The sampling process is governed by a distribution $P_{\nu}$, which is the joint distribution over the inclusion indicator variables $(\delta_i, \ldots \delta_N)$. While the analyst is only interested in making inference about the population model  $P_{\theta_{0}}$, they must deal with the joint distribution $P_{\theta_{0}}, P_{\nu}$ which governs the observed sample they have at hand.

$P_{\nu}$ defines the marginal probabilities of selection (or inclusion) $\pi_i = Pr(\delta_i = 1)$ as well as all higher order joint inclusions, e.g. $\pi_{ij} = Pr(\delta_i = 1, \delta_j = 1)$. These probabilities are directly set or induced by decisions from the study designers. For example, $\pi_i$ are often chosen to be associated with the variable of interest to increase the efficiency of resulting summary statistics such as totals or means. Such a design $P_{\nu}$ is called `informative', because the balance of information in the sample is different from the balance of information in the original population. 

Informative sample designs pose two challenges for inference: (i) Estimating $P_{\theta_{0}}$ on the sample without accounting for unequal selection  $\pi_i$ will lead to biased estimates. (ii) Failure to account for dependence induced by $P_{\nu}$ (for example through $\pi_{ij}$) will lead to models with misspecified dependence structures, which leads to uncertainty estimates such as posterior intervals that are asymptotically incorrect.

The first challenge can be addressed by using the survey weights $w_i \propto 1/\pi_i$. In order to approximate the population posterior
\begin{gather*}
    \pi\left(\bm{\theta}\vert \mathbf{y}\right) \propto \mathop{\prod}_{i = 1}^{N}\pi\left(y_{i}\vert \bm{\theta}\right)\pi\left(\bm{\theta}\right),
    \end{gather*}

we use the survey-weighted pseudo-posterior for the observed sample
\begin{gather*}
    \pi^{\pi}\left(\bm{\theta}\vert \mathbf{y},\mathbf{w}\right) \propto \textcolor{red}{\left[\mathop{\prod}_{i = 1}^{n}\pi\left(y_{i}\vert \bm{\theta}\right)^{w_{i}}\right]}\pi\left(\bm{\theta}\right).
    \end{gather*}

This is a relatively minor change to the underlying model and easy to implement, especially in general modelling software which already allows for weights. As we will see in our examples (Section \ref{example-3}), {\bf brms} has a key word `weights' associated with the dependent variable when specifying the form of the model via \texttt{brmsformula}.
This modification leads to consistent estimation (i.e. the collapsing of the posterior around the true value $\theta_0$). This consistency is with respect to the joint process of the population generation $P_{\theta_{0}}$ and the taking of samples $P_{\nu}$ (for more technical details, see  \cite{2015arXiv150707050S} and \cite{2018dep}). It is worth noting that the consistency for the survey-weighted pseudo maximum likelihood estimator (MLE) has been known for many years and has been demonstrated for multiple applications. See \cite{Isaki82} for an example of an earlier reference and \cite{Han2021} for a recent comprehensive treatment.

The second challenge is asymptotically correct uncertainty quantification. The survey-weighted pseudo-posterior is not `fully' Bayesian, in that it is not fully generative. The resulting estimated model and parameters \emph{can} be used to generate population level outcomes, but not the original sample. Thus these models are `partially' generative, because they condition on the observed sample selection $P_{\nu}$ but do not model the joint process $P_{\theta_{0}},P_{\nu}$. While consistent estimation for $P_{\theta_{0}}$ is possible, the models are misspecified with respect to $P_{\nu}$. These kinds of `working' models lead to incorrect posterior intervals, in the sense that their asymptotic frequentest coverages are incorrect (e.g. a 95\% posterior interval does not have 95\% coverage). See \cite{2009arXiv0911.5357R} for a closely related example for posterior inference based on composite likelihoods. For fully generative models, we expect asymptotically that the MLE and the posterior distribution will be normal with the same mean and variance $N(\theta_0, H^{-1}_{\theta_0})$ (called a Bernstein-von Mises result). For a textbook level presentation of Bernstein-von Mises results for many classes of Bayesian models, see \cite{ghosal2017fundamentals}. For misspecified or `working' models such as the survey-weighted pseudo-posterior, the two will have different covariance matrices. See \cite{kleijn2012} for more technical details. 

To facilitate comparisons between covariance matrices, we first use the following empirical distribution approximation for the joint distribution over population generation and the draw of an informative sample that produces the observed sample.  This empirical distribution construction follows \citet{breslow:2007} and incorporates inverse inclusion probability weights, $\{1/\pi_{i}\}_{i=1,\ldots,N}$, to account for the informative sampling design \citep{2015arXiv150707050S}.
\begin{equation}
\mathbb{P}^{\pi}_{N_{\nu}} = \frac{1}{N}\mathop{\sum}_{i\in U}\frac{\delta_{i}}{\pi_{ i}}\delta\left(y_{i}\right),
\end{equation}
where $\delta\left(y_{i}\right)$ denotes the Dirac delta function, with probability mass $1$ on $y_{i}$. This construction contrasts with the usual empirical distribution, $\mathbb{P}_{N} = \frac{1}{N}\mathop{\sum}_{i\in U}\delta\left(y_{i}\right)$. Using the notation of \citet{Ghosal00convergencerates} we define expectation functionals with respect to these empirical distributions by $\mathbb{P}^{\pi}_{N}f = \frac{1}{N}\mathop{\sum}_{i=1}^{N}\frac{\delta_{i}}{\pi_{i}}f\left(y_{ i}\right)$.  Similarly, $\mathbb{P}_{N}f = \frac{1}{N}\mathop{\sum}_{i=1}^{N}f\left(y_{i}\right)$.

The asymptotic covariances are then 
\begin{itemize}
    \item Simple Random Sample MLE: $H^{-1}_{\theta_0}$
    \item Survey-weighted MLE:  $H^{-1}_{\theta_0}\textcolor{red}{J^{\pi}_{\theta_0}}H^{-1}_{\theta_0}$
    \item Survey-weighted Posterior: $H^{-1}_{\theta_0}$
\end{itemize}
where 
\begin{equation*}
H_{\theta_{0}} = - \mathbb{E}_{P_{\theta_{0}}} \left[\mathbb{P}_{N} \ddot{\ell}_{\theta_{0}}\right]
= -\frac{1}{N}\mathop{\sum}_{i\in U}\mathbb{E}_{P_{\theta_{0}}}\ddot{\ell}_{\theta_{0}}(y_{i}),
\end{equation*} 
is the Fisher Information, i.e. the negative of the expected value of the second derivative of the log-likelihood $\ddot{\ell}_{\theta_{0}}$. We see that the survey-weighted version based on the sample  $H^{\pi}_{\theta_{0}} =  H_{\theta_{0}}$:
\begin{equation*}
 H^{\pi}_{\theta_{0}} = - \mathbb{E}_{P_{\theta_{0}}, P_{\nu}} \left[\mathbb{P}^{\pi}_{N_{\nu}} \ddot{\ell}_{\theta_{0}}\right]
 = -\frac{1}{N}\mathop{\sum}_{i\in U}\mathbb{E}_{P_{\theta_{0}}}\left[\mathbb{E}_{P_{\nu}}\frac{\delta_{ i}}{\pi_{ i}}\ddot{\ell}_{\theta_{0}}(y_i)\right] = -\frac{1}{N}\mathop{\sum}_{i\in U}\mathbb{E}_{P_{\theta_{0}}}\ddot{\ell}_{\theta_{0}}(y_{i}).
\end{equation*}

\begin{figure}
\centering
\includegraphics[width = 0.60\textwidth,
		page = 1,clip = true, trim = 0.25in 0.25in 0in 0.in]{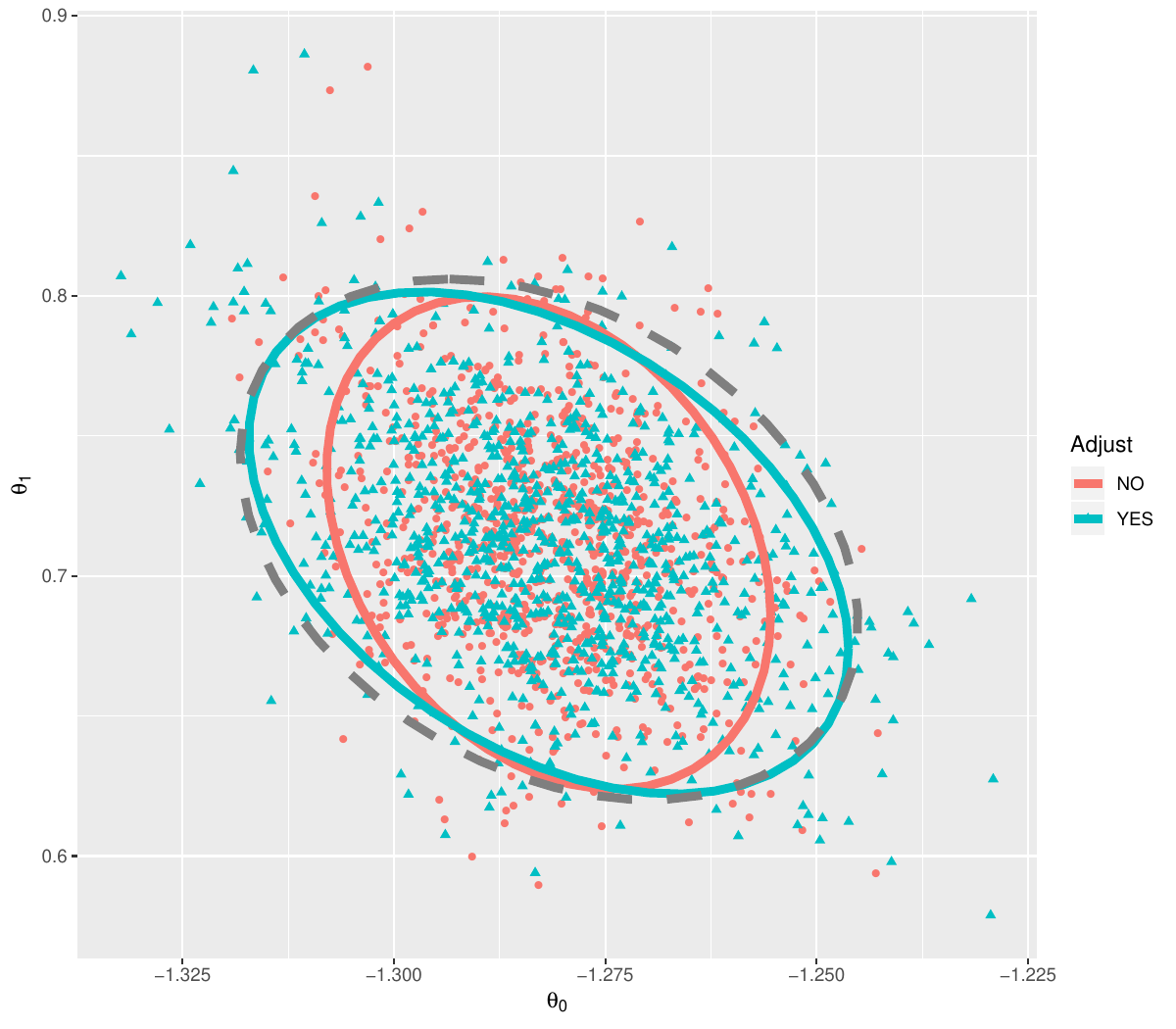}
\caption{Joint pseudo-posterior sample for the intercept (horizontal) and slope (vertical) for a logistic regression modeling current cigarette smoking by past year major depressive episode based the 2014 National Survey on Drug Use and Health. Unadjusted (red circles) and adjusted (blue triangles) draws with approximate 90\% density ellipses. Asymptotic normal 90\% ellipse for pseudo-MLE (dashed). Reproduced from \cite{WilliamsISR21}.}
\label{fig:NSDUHscatter}
\end{figure}

The survey-weighted posterior behaves as if it were taken from a simple random sample. In contrast, the survey-weighted MLE has a sandwich form to its variance, where $J^{\pi}_{\theta_0}$ is the variance of the gradient of the log-likelihood with respect to the joint distribution of $P_{\theta_{0}},P_{\nu}$.
\begin{equation*}
\begin{array}{rl}
J^{\pi}_{\theta_{0}} \ = & 
\mathbb{V}_{P_{\theta_{0}},P_{\nu}}\left[\mathbb{P}^{\pi}_{N_{\nu}}\dot{\ell}_{\theta_{0}} \right]
=\mathbb{E}_{P_{\theta_{0}},P_{\nu}}\left[\left(\mathbb{P}^{\pi}_{N_{\nu}}\dot{\ell}_{\theta_{0}}\right)\left(\mathbb{P}^{\pi}_{N_{\nu}}\dot{\ell}_{\theta_{0}}\right)^{T} \right] 
\\
J_{\theta_{0}} \ = &
\mathbb{V}_{P_{\theta_{0}}}\left[\mathbb{P}_{N}\dot{\ell}_{\theta_{0}} \right] = 
\mathbb{E}_{P_{\theta_{0}}}\left[\left(\mathbb{P}_{N}\dot{\ell}_{\theta_{0}}\right)\left(\mathbb{P}_{N}\dot{\ell}_{\theta_{0}}\right)^{T} \right] 
\\
J^{\pi}_{\theta_{0}} \ \ne & J_{\theta_{0}}
\end{array}
\end{equation*}
Under correctly specified models, $J_{\theta_0} = H_{\theta_0}$ (Bartlett's $2^{nd}$ identity) and the sandwich collapses. However for misspecified models, such as those using a pseudo-likelihood, these two matrices are distinct $J^{\pi}_{\theta_{0}} \ne H_{\theta_{0}}$, leading to the sandwich form (Godambe Information matrix). Figure \ref{fig:NSDUHscatter} compares the bi-variate posterior for a slope and intercept logistic regression model from a complex household survey. Compared to the asymptotic distribution of the MLE, the unadjusted pseudo-posterior has incorrect scale and rotation. However an adjustment leads to a posterior that more closely aligns with the MLE. The next section provides details on the formulation and implementation of the adjustment.

\hypertarget{adjusting the-survey-weighted-pseudo-posterior}{%
\subsection{Adjusting the Survey-Weighted Pseudo-Posterior}\label{adjusting the-survey-weighted-pseudo-posterior}}
Similar to \cite{2009arXiv0911.5357R}, we will use a curvature adjustment to the pseudo-posterior sample. However, following \cite{WilliamsISR21} we perform this adjustment post-hoc, instead of inserting significant complexities into the underlying MCMC sampling process. Let $\hat{\theta}_m$ represent the sample from the pseudo-posterior for $m = 1, \ldots, M$ draws with sample mean $\bar{\theta}$.
Define the adjusted sample:
\begin{equation}\label{eq:adjustment}
\hat{\theta}^{a}_m = \left(\hat{\theta}_m - \bar{\theta}\right) R^{-1}_2 R_1  + \bar{\theta},
\end{equation}
where $R_1$ and $R_2$ are `square root' matrices such that $R'_1 R_1 = H_{\theta_{0}}^{-1} J^{\pi}_{\theta_{0}} H_{\theta_{0}}^{-1}$ and $R'_2 R_2 = H_{\theta_{0}}^{-1}$.
Asymptotically, $\hat{\theta}_m \appsim N(\theta_{0}, n^{-1} H_{\theta_{0}}^{-1})$. Then we now have $\hat{\theta}^{a}_m \appsim N(\theta_{0}, n^{-1} H_{\theta_{0}}^{-1} J^{\pi}_{\theta_{0}} H_{\theta_{0}}^{-1})$, which is the asymptotic distribution of the MLE under the pseudo-likelihood. In survey statistics, we often estimate a `design effect' \citep[see, for example][]{kish1995methods} which is a ratio of the (asymptotic) variance of an estimate under the complex survey design vs. its (asymptotic) variance under a simple random sample. This helps us understand the penalty or efficiency of the design (in terms of variance per fixed sample size) vs. a simpler design. In this way, the adjustment $R^{-1}_2 R_1$ is a multivariate estimate of a design effect, providing a parameter-specific adjustment for effective sample size for variances and as well as for correlations between parameters.

For pseudo-posteriors using working likelihoods such as a marginal composite likelihoods \citep{2009arXiv0911.5357R} or an asymmetric Laplace distribution for quantile regression \citep{insr.12114}, the form of  $\mbox{Var}_{P_{\theta_{0}}}\dot{\ell}_{\theta_{0}} =  J_{\theta_{0}}$ may be derived analytically. 
However, for the survey-weighted pseudo-posterior, the additional sampling design $P_{\nu}$ is unlikely to be in an analytic form. In rare cases, the full $n \times n$ matrix of pairwise inclusion probabilities $\pi_{ij}$ may be available or approximated. In that case, an expression for variance of a weighted sum such as the score function $\dot{\ell}_{\theta_{0}}$ is readily available and could be used for estimating $J^{\pi}_{\theta_{0}}$ \citep[See, for example][]{Wang2017}. In practice, the design is often algorithmically defined; for example designs may use the sorting and clustering of population units in addition to unequal probabilities of selection. Rather than assuming a simplifying model for this distribution (e.g. $\pi_i \propto y_i$), we instead approximate the joint distribution $(P_{\theta_{0}}, P_{\nu})$ with the empirical distribution by re-sampling the units and associated response values. In the next section, we describe how this approaches fits in with the tradition of variance estimation in survey statistics.

\hypertarget{variance-estimation-in-complex-samples}{%
\subsection{Variance Estimation in Complex Samples}\label{variance-estimation-in-complex-samples}}
The de-facto approach for variance estimation implemented in survey analysis packages starts with the assumption of approximate independence of the sampling of the first stage unit or primary clusters, sometimes called `ultimate' clusters \citep{kalton1979ultimate, heeringa2010applied}. The idea is that any subsequent within-cluster dependence based on additional stages of sampling is a nuisance that is removed by aggregating to the primary sampling unit (PSU), much like repeated measurements of an individual can be aggregated up to a individual average, with variance estimation based on differences \emph{between} averages of individuals. We note that consistency of estimation is still possible with the survey-weighted pseudo-posterior even in the presence of these un-modelled dependencies within clusters \citep{2018dep}. The two major classes of estimation approaches under this framework are Taylor linearization (sometimes called infintessimal jackknife) and replication methods. We describe each at a high level but point the interested reader to the rich literature on comparisons and variations of these approaches applied to survey-weighted estimating functions \citep[See for example][as a starting point]{binder1996, Rao92}. 

Let $y_{ij}$, and $w_{ij}$ be the observed data  and sampling weights for individual $i$ in cluster $j$ of the sample. Assume the parameter $\theta$ is a vector of dimension $d$ with population model value $\theta_0$.
The main idea behind the Taylor linearization approach is to approximate the point estimate $\hat{\theta}$, or a centered `residual' version $(\hat{\theta} - \theta_{0})$, as a simple weighted sum:  $\hat{\theta} \approx \sum_{i,j} w_{ij} z_{ij}(\theta)$. The function $z_{ij}$ is the first order derivative evaluated at the current values of $y_{ij}$, and $\hat{\theta}$. In this way the approximation for $\hat{\theta}$ is the simple sum across independent PSU's: $\hat{\theta} \approx \sum_{j} \hat{\theta}_j$ where each is approximated by the sum $\hat{\theta}_j = \sum_{i} w_{ij} z_{ij} (\theta)$. Then the variance estimate of $\hat{\theta}$ is simply taken as the between PSU variance of the $\hat{\theta}_j$: $\widehat{Var(\hat{\theta})} = \frac{1}{J-d}\sum_{j =1}^{J}(\hat{\theta} -\hat{\theta}_j )(\hat{\theta} -\hat{\theta}_j )^{T}$. For stratified sampling designs, these variances are calculated separately within strata and summed.

For a replication-based method, we can use randomization (e.g. bootstrap), leave-one-out (e.g. jackknife), or efficient orthogonal contrasts (balanced repeated replication) to create a set of $K$ replicate weights $(w_{i})_k$ each of the same size as our original sample weights (e.g. for all $i \in S$ and for every $k = 1,\ldots, K$). Typically each replicate weight has a modified values (often 0) for a subset of clusters. Other clusters often receive an increase in weight to compensate so that the sum of the weights are consistent with the original weights. For each replicate $k\in 1,\ldots, K$, we repeat the complete estimation process to produce $\hat{\theta}_k$. For a Bayesian analysis, this would typically mean running a new MCMC sampler for each $k$ replicate. We then compute the simple variance between replicates $\widehat{Var(\hat{\theta})} = \frac{C}{K-d}\sum_{k =1}^{K}(\hat{\theta} -\hat{\theta}_k )(\hat{\theta} -\hat{\theta}_k )^{T}$ where a normalizing or multiplicity constant $C$ may be needed depending on the method (e.g. jackknife replications). For stratified samples, replicates are generated such that each strata is represented in every replicate.

For the purposes of applying these variance estimation approaches to parameter estimates from more complex Bayesian models, each of these methods present challenges. In terms of \emph{computational effort}, the Taylor linearization method only requires posterior estimation one time, for the full sample. Variance estimates then use the posterior mean or another plug-in value (e.g. median or mode). In contrast, the replication method requires a separate posterior estimation for \emph{each} of the $K$ replicates, where $K$ is commonly on the order of 100 for national studies. In terms of analytical challenges, the Taylor linearization approach requires the calculation of the gradient of the log-likelihood as part of $z_{ij} (\theta)$. This requires significant \emph{analytic challenges} for all but the simplest models. Current survey software has these derivatives hard-coded. Thus expanding a new set of models requires the user to provide the functional form of these derivatives. See for example the \texttt{svymle} function of the {\bf survey} package \citep{Rsurvey}.

\hypertarget{introducing-the-cssampling-package}{%
\subsection{Introducing the csSampling package}\label{introducing-the-cssampling-package}}

The {\bf csSampling} package uses existing R packages to address the challenges of these two variance estimation methods. We first note that recent advances in algorithmic differentiation \citep{Margossian18} allow us to specify a model as a log density but only treat its gradient in the abstract \emph{without} specifying it analytically. This approach is a core part of the Stan software and the gradient of the log density of a Stan model is exposed to R through \texttt{grad\_log\_prob} of the {\bf rstan} package \citep{Rstan}.

We first use the unadjusted survey-weighted posterior to estimate the Fisher Information $H_{\theta_{0}}$.
We recall from Section \ref{inference-from-complex-survey-samples} that $H_{\theta_{0}} = -\mathbb{E}_{P_{\theta_{0}}}\ddot{\ell}_{\theta_{0}}$ and $ H^{\pi}_{\theta_{0}} =-\mathbb{E}_{P_{\theta_{0}},P_{\nu}}\left[\mathbb{P}^{\pi}_{N_{\nu}}\ddot{\ell}_{\theta_{0}}\right] = H_{\theta_{0}}$.
Therefore, consistent estimates of $H_{\theta_{0}}$ are available using either the plug-in estimate $-\mathop{\sum}_{i \in S} w_i \ddot{\ell}_{\bar{\theta}}(y_{i})$ using the posterior mean $\bar{\theta}$ or the average over posterior samples $-\frac{1}{M}\mathop{\sum}_{m=1}^{M} \mathop{\sum}_{i \in S} w_i \ddot{\ell}_{\hat{\theta}_{m}}(y_{i})$. We note that the former is faster while the latter is more stable (more likely to be positive definite) which tends to help for more complicated models. By default, \texttt{cs\_sampling} uses the latter Monte Carlo average, but the plug-in approach is also available. We can obtain $-\mathop{\sum}_{i \in S} w_i \ddot{\ell}_{\theta}(y_{i})$ as a function of $\theta$ by using the gradient from \texttt{grad\_log\_prob} and finite difference approaches via \texttt{optimHess} in the {\bf stats} package in R.

As noted in Section \ref{inference-from-complex-survey-samples},  $J^{\pi}_{\theta_{0}}  \ne J_{\theta_{0}}$. The weighted pseudo-posterior samples alone are not sufficient to estimate $J^{\pi}_{\theta_{0}}$.
We use a hybrid approach to variance estimation that allows for Taylor linearization for replication designs. Instead of directly estimating the variance of $\hat{\theta}$, we are interested in the variance of the transformed parameters:
\begin{equation}\label{eq:variance}
(\hat{\psi} - \psi_0) = H_{\theta_0} (\hat{\theta} - \theta_0) \approx \mathop{\sum}_{i \in S} w_i \dot{\ell}_{\hat{\theta}}(y_{i}) = \mathop{\sum}_{i \in S} w_i z_i (\hat{\theta}),
\end{equation}
where $Var_{P_{\theta_0}, P_{\nu}} (\hat{\psi} - \psi_0) \approx J^{\pi}_{\theta_{0}}$.
In other words, in order to estimate $J^{\pi}_{\theta_{0}}$, we use a variance estimation method for the linearized $\mathop{\sum}_{i \in S} w_i \dot{\ell}_{\hat{\theta}}(y_{i})$ using a plug-in estimate $\hat{\theta}$. We choose to use replication methods to perform this estimation, because many studies only provide replication weights, while studies that use cluster and strata identification can be readily converted into replication designs. For example, we will use the \texttt{as.svrepdesign} of the {\bf survey} package \citep{Rsurvey} which allows the users to create a variety of replication weights based on strata and cluster information. \cite{WilliamsISR21} originally proposed an approach of creating bootstrap replicate weights based on resampling half the PSUs in each strata. This approach is readily available in the {\bf survey} package and is the default when using {\bf csSampling}.

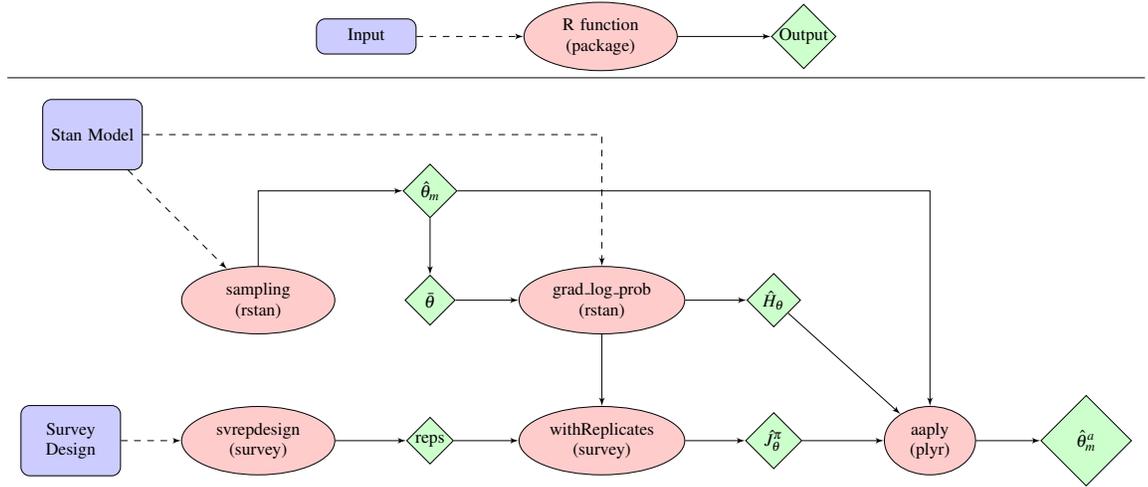
\begin{figure}
\centering
\tikzstyle{decision} = [diamond, draw, fill=green!20, text width=4em, text badly centered, node distance=2.5cm, inner sep=0pt]
\tikzstyle{decision2} = [diamond, draw, fill=green!20, text badly centered, node distance=2.25cm, inner sep=0pt]
\tikzstyle{decision3} = [diamond, draw, fill=green!20, text badly centered, node distance=3.25cm, inner sep=0pt]
\tikzstyle{smdecision} = [diamond, draw, fill=green!20, text width=2em, text badly centered, node distance=2.75cm, inner sep=0pt]
\tikzstyle{smdecision2} = [diamond, draw, fill=green!20, text width=2em, text badly centered, node distance=1.75cm, inner sep=0pt]
\tikzstyle{block} = [rectangle, draw, fill=blue!20, text width=5em, text centered, rounded corners, minimum height=4em, node distance = 3cm]
\tikzstyle{block2} = [rectangle, draw, fill=blue!20, text centered, text width=5em, minimum height=2em, rounded corners, node distance = 3.75cm]
\tikzstyle{line} = [draw, -latex']
\tikzstyle{cloud} = [draw, ellipse, fill=red!20, node distance=3.75cm, minimum height=2em,text centered, text width=5.5em]
\tikzstyle{clouda} = [draw, ellipse, fill=red!20, node distance=2.25cm, minimum height=2em,text centered, text width=5.5em]
\tikzstyle{cloud2} = [draw, ellipse, fill=red!20, node distance=2.75cm, minimum height=2em,text centered, text width=6em]
\tikzstyle{cloud3} = [draw, ellipse, fill=red!20, node distance=2.5cm, minimum height=2em,text centered, text width=3em]
\tikzstyle{cloud4} = [draw, ellipse, fill=red!20, node distance=3.75cm, minimum height=2em,text centered, text width=5.5em]

\resizebox{6.0in}{!}{
\begin{tabular}{c}
\begin{tikzpicture}[node distance = 2cm, auto,]
\node [cloud4] (r1) {R function (package)};
\node [block2, left of = r1 ] (i1) {Input};
\node [decision3, right of = r1] (d1) {Output};
\path [line, dashed] (i1) -- (r1);
\path [line] (r1) -- (d1);
\end{tikzpicture}
\\\hline\\
\begin{tikzpicture}[node distance = 2cm, auto,]
\node [block] (stan) {Stan Model};
\node [cloud, below right of=stan] (rstan) {sampling (rstan)};
\node [clouda, below of=rstan] (rsurvey) {svrepdesign (survey)};
\node [block, left of=rsurvey] (survey) {Survey Design};
\node [smdecision, right of=rsurvey] (reps) {reps};
\node [smdecision, right of=rstan] (thetab) {$\bar{\theta}$ };
\node [cloud2, right of=thetab] (rstan2) {grad\_log\_prob (rstan)};
\node [smdecision2, above of = thetab] (thetam) {$\hat{\theta}_m$};
\node [smdecision, right of = rstan2] (H) {$\hat{H}_{\theta}$};
\node [cloud2, right of=reps] (rsurvey2) {withReplicates (survey)};
\node [smdecision, right of = rsurvey2] (J) {$\hat{J}^{\pi}_{\theta}$};
\node [cloud3, right of=J] (aaply) {aaply (plyr)};
\node [decision, right of=aaply] (thetaadj) {$\hat{\theta}^{a}_m$};
\path [line, dashed] (survey) -- (rsurvey);
\path [line] (rsurvey) -- (reps);
\path [line] (reps) -- (rsurvey2);
\path [line] (rsurvey2) -- (J);
\path [line] (J) -- (aaply);
\path [line] (aaply) -- (thetaadj);
\path [line, dashed] (stan) -- (rstan);
\path [line] (rstan) |- (thetam);
\path [line] (thetam) -- (thetab);
\path [line] (thetab) -- (rstan2);
\path [line] (rstan2) -- (rsurvey2);
\path [line] (rstan2) -- (H);
\path [line] (H) -- (aaply);
\path [line] (thetam) -| (aaply);
\path [line, dashed] (stan) -| (rstan2);
\end{tikzpicture}

\end{tabular}
}
\caption{Flowchart of the implementation of estimation and adjustment in {\bf csSampling}.}
\label{fig:flowchart}
\end{figure}

Once we have an estimate of $J^{\pi}_{\theta_{0}}$, we can combine it with an estimate of $H_{\theta_{0}}$. We then calculate the `square root' matrices $R_1$ and $R_2$ and apply the adjustment from Equation \ref{eq:adjustment}. Figure \ref{fig:flowchart} presents a flowchart of the {\bf csSampling} algorithm.  

\hypertarget{comparison-to-other-packages}{%
\subsection{Comparison to Other Packages}\label{comparison-to-other-packages}}
The {\bf csSampling} package extends the functionality of the {\bf brms} and {\bf survey} packages in R. By itself {\bf brms} can be used to estimate survey-weighted pseudo-posterior distributions, which are asymptotically consistent. Other packages that allow for weighted likelihoods can also be used, for example {\bf rstanarm}  \citep{rstanarm}. These packages could also be used with the \texttt{withReplicates} function from the {\bf survey} package. This would be a pure replication approach, requiring a new posterior estimation for each replication (e.g. 100 separate runs). This approach could be used to create a design effect by estimating the current posterior variance $V_P (\bar{\theta})$ and the replication based variance $V_R (\bar{\theta})$. Asymptotically $V_P \rightarrow H^{-1}_{\theta_0}$ and $V_R \rightarrow H^{-1}_{\theta_0} J^{\pi}_{\theta_{0}} H^{-1}_{\theta_0}$. So these matrices could also be used to generate $R_2$ and $R_1$. The {\bf csSampling} exploits the use of {\bf rstan} to provide the gradients of the log likelihood, allowing the use of a plug-in approach for hybrid replication estimation which eliminates the need for separate posterior estimation for each replication.

Similarly, the \texttt{surveymle} from  the {\bf survey} package can take general functions for likelihoods and their gradients, for example those provided by \texttt{grad\_log\_prob} from {\bf rstan}, and produce survey-weighted MLE point $\theta_{MLE}$ and asymptotic variance estimates based on the pure Taylor linearization approach $V_{TL} (\theta_{MLE}) = H^{-1}_{\theta_{MLE}} J^{\pi}_{\theta_{MLE}} H^{-1}_{\theta_{MLE}}$. The current approach in {\bf csSampling} provides a survey-weighted (pseudo-) Bayesian approach. The advantage of the latter is that small sample properties of the posterior are maintained, with the adjustment allowing for better asymptotic properties.

\hypertarget{the-package}{%
\section{The package}\label{the-package}}

The \textbf{csSampling} package is available on GitHub and can be obtained in R using the following commands: 

\begin{lstlisting}

    library(devtools)
    install_github("RyanHornby/csSampling")
    library(csSampling)
    rstan_options(auto_write = TRUE)
    
    \end{lstlisting}

The {\bf csSampling} depends on {\bf rstan} \citep{Rstan}, {\bf brms} \citep{brms} and {\bf survey} \citep{Rsurvey} R packages. Since {\bf rstan} compiles Stan models into C++ code, setting the \texttt{rstan\_options} will save the compiled object and avoid extra compiling across sessions.

The {\bf csSampling} package provides a main function, called \texttt{cs\_sampling}, and two auxiliary functions, called \texttt{cs\_sampling\_brms} and \texttt{plot}. There are other functions that are used to facilitate compiling the codes in the main and auxiliary functions. However those helper functions are not released for data users (i.e., \texttt{DEadj}, \texttt{grad\_par} and \texttt{list\_2D\_row\_subset}).

\hypertarget{the-main-function}{%
\subsection{The main function}\label{the-main-function}}

The \texttt{cs\_sampling} is a wrapper function that accommodates the following inputs: a \texttt{svydesign} object (of the {\bf survey} package), a \texttt{stan\_model} (of the {\bf rstan}), and inputs for \texttt{sampling} (of the {\bf rstan}). The \texttt{cs\_sampling} calls the \texttt{sampling} to generate MCMC draws from the model. The constrained parameters are converted to unconstrained, adjusted, converted back to constrained and then output. The \texttt{cs\_sampling} signature is:

\begin{lstlisting}

    cs_sampling(svydes, mod_stan, par_stan, data_stan,
                ctrl_stan, rep_design, ctrl_rep, 
                H_estimate, matrix_sqrt, sampling_args)
    
    \end{lstlisting}

The function arguments are defined as below:

    \begin{itemize}
        \item \texttt{svydes}: an object of complex survey information defined by \texttt{svydesign} or \texttt{svrepdesign} of the {\bf survey} package.
        \item \texttt{mod\_stan}: a compiled Stan model from \texttt{stan\_model}.
        \item \texttt{par\_stan}: a list of a subset of parameters to output after adjustment. All parameters are adjusted including the derived parameters, so users may want to only compare subsets. The default, NA, will return all parameters.
        \item \texttt{data\_stan}: a list of data inputs for sampling associated with \texttt{mod\_stan}.
        \item \texttt{ctrl\_stan}: a list of control parameters to pass to sampling. Parameters includes the number of chains, iterations, warmup, and thin with defaults "\texttt{list(chains = 1, iter = 2000, warmup = 1000, thin = 1)"}.
        \item \texttt{rep\_design}: logical indicating if the \texttt{svydes} object is a \texttt{svrepdesign}. If FALSE, the design will be converted to a \texttt{svrepdesign} using \texttt{ctrl\_rep} settings.
        \item \texttt{ctrl\_rep}: a list of settings when converting \texttt{svydes} from a \texttt{svydesign} object to a \texttt{svrepdesign} object. Settings include \texttt{replicates}, number of replicate weights, and \texttt{type}, the type of replicate method to use. The default is "\texttt{list(replicates = 100, type = "mrbbootstrap")}".
        \item \texttt{H\_estimate}: a string indicating the method to use to estimate H. The default "MCMC" is Monte Carlo averaging over posterior draws. Otherwise, a plug-in using the posterior mean.
        \item \texttt{matrix\_sqrt}: a string indicating the method to use to take the "square root" of the R1 and R2 matrices. The default "eigen" uses the eigenvalue decomposition. Otherwise, the Cholesky decomposition is used.
        \item \texttt{sampling\_args}: a list of extra arguments that get passed to sampling.
    \end{itemize}

\hypertarget{auxiliary-functions}{%
\subsection{Auxiliary functions}\label{auxiliary-functions}}
The \texttt{cs\_sampling\_brms} is a wrapper that uses helper functions from {\bf brms} to create stan code and data inputs that are then passed to \texttt{cs\_sampling}.
The \texttt{cs\_sampling\_brms} signature is:

\begin{lstlisting}

    cs_sampling_brms(svydes, brmsmod, data, family, par_brms,
                    prior, stanvars, knots, ctrl_stan, rep_design,
                    ctrl_rep, stancode_args, standata_args, 
                    H_estimate, matrix_sqrt, sampling_args)
    
    \end{lstlisting}

The function arguments are defined as below:

    \begin{itemize}
        \item \texttt{svydes}: an object of complex survey information defined by \texttt{svydesign} or \texttt{svrepdesign} of the {\bf survey} package.
        \item \texttt{brmsmod}: a \texttt{brmsformula} object that is an input of \texttt{make\_stancode}.
        \item \texttt{data}:  data frame, as input to \texttt{make\_stancode}.
        \item \texttt{family}: a \texttt{brmsfamily} as input to \texttt{make\_stancode} specifying distribution and link function.
        \item \texttt{par\_brms}: a list of a subset of parameters to output after adjustment. All parameters are adjusted including the derived parameters, so users may want to only compare subsets. The default, NA, will return all parameters.
        \item \texttt{prior}: optional input to \texttt{make\_stancode}.
        \item \texttt{stanvars}: optional input to \texttt{make\_stancode}.
        \item \texttt{knots}: optional input to \texttt{make\_stancode}.
        \item \texttt{ctrl\_stan}: a list of control parameters to pass to \texttt{sampling}. Currently includes the number of chains, iterations, warmpup, and thin.  The default is "\texttt{list(chains = 1, iter = 2000, warmup = 1000, thin = 1)}".
        \item \texttt{rep\_design}: logical indicating if the \texttt{svydes} object is a \texttt{svrepdesign}. If FALSE, the design will be converted to a \texttt{svrepdesign} using \texttt{ctrl\_rep} settings.
        \item \texttt{ctrl\_rep}: a list of settings when converting \texttt{svydes} from a \texttt{svydesign} object to a \texttt{svrepdesign} object. Currently includes the number of replicates and the type of replication method. The default is "\texttt{list(replicates = 100, type = "mrbbootstrap")}".
        \item \texttt{stancode\_args}: a list of extra arguments to be passed to \texttt{make\_stancode}.
        \item \texttt{standata\_args}: a list of extra arguments to be passed to \texttt{make\_standata}.
        \item \texttt{H\_estimate}: a string indicating the method to use to estimate H. The default "MCMC" is Monte Carlo averaging over posterior draws. Otherwise, a plug-in using the posterior mean.
        \item \texttt{matrix\_sqrt}: a string indicating the method to use to take the "square root" of the R1 and R2 matrices. The default "eigen" uses the eigenvalue decomposition. Otherwise, the Cholesky decomposition is used.
        \item \texttt{sampling}: a list of extra arguments to be passed to \texttt{sampling}.
    \end{itemize}

The \texttt{plot} signature is:

\begin{lstlisting}

    plot(x, varnames, ...)
    
    \end{lstlisting}

The function arguments are defined as below:

    \begin{itemize}
        \item \texttt{x}: a \texttt{cs\_sampling} output.
        \item \texttt{varnames}: optional vector of names of a subset of variables for pairs plotting.
    \end{itemize}

\hypertarget{illustrations}{%
\section{Illustrations}\label{illustrations}}

In this section, we present three examples where {\bf csSampling} is used to model different types of dependent variables, such as continuous, multinomial and Bernoulli-distributed variables. In each example, we highlight different features of {\bf csSampling}.

\hypertarget{example-1}{%
\subsection{Example 1: continuous dependent variable
}\label{example-1}}

In this example, \texttt{apistrat} of the \texttt{api} data sets from the {\bf survey} R package is used. The \texttt{api} data sets contain various probability samples of data from California schools with at least 100 students. The \texttt{apistrat} is a stratified sample of 200 schools with data about student performance. The sample is stratified according to the school type, \texttt{stype}, with 100 elementary schools, 50 middle schools, and 50 high schools. In this example, we highlight the use of the \texttt{cs\_sampling\_brms} function of {\bf csSampling} to model \texttt{api}, i.e., Academic Performance Index (API) in 2000, with the following variables as dependent variables: 1) \texttt{ell}: percentage of students who are English Language Learners, 2) \texttt{meals}: percentage of students eligible for subsidized meals, and 3) \texttt{mobility}: percentage of students for whom this is the first year at the school. The population model is a simple linear regression:
\[
\begin{array}{rl}
    y_i|\mu_i, \sigma &  \sim \ N(\mu_i, \sigma) \\
    \mu_i & = \  X_i \beta \\
    \beta & \sim \ \propto 1 \\
    \sigma & \sim t_{\nu}(0, \phi)
\end{array}
\]
where we have a flat (improper prior) for regression coefficients $\beta$ and a (truncated) student t distribution for $\sigma$ with degrees of freedom $\nu$ and scale $\phi$. These are the defaults used by {\bf brms}, but the user can also provide other priors to pass to {\bf brms}.

To estimate this population level model from the survey data, we run a survey-weighted linear regression following the steps below: 
\begin{enumerate}
  \item Define data and parameters of the complex sample: After reading the data, \texttt{svydesign} is used to define the complex sample parameters, i.e., strata (stype), weights (pw) and finite population correction (fpc). The survey weights \texttt{wt} should first be re-scaled, i.e., normalized to sum to the sample size, as required to match those used for the Stan model.

\begin{lstlisting}
    data(api)
    apistrat$wt <- apistrat$pw/mean(apistrat$pw)
    dstrat <- svydesign(id=~1,
    +                        strata= ~stype, 
    +                        weights= ~wt, 
    +                        data= apistrat, 
    +                        fpc= ~fpc)
\end{lstlisting}

  \item Define and run the Stan model Via BRMS: We start the process by running a Stan model using the \texttt{cs\_sampling\_brms} function after accounting for the complex design features defined in \texttt{dstrat}. 
    
\begin{lstlisting}

    set.seed(12345)
    model_formula <- formula("api00|weights(wt) 
    +                           ~ ell + meals + mobility")
    mod.brms <- cs_sampling_brms(svydes = dstrat,
    +                brmsmod = brmsformula(model_formula, center = FALSE), 
    +                data = apistrat, 
    +                family = gaussian())
    
    \end{lstlisting}
    
The wrapper {\bf cs\_sampling\_brms} calls the {\bf brms} helper function \texttt{make\_stancode} whichs generates the following Stan code for a survey-weighted linear regression model. The code is then compiled and passed to \texttt{cs\_sampling}.
\begin{lstlisting}
   
    // generated with brms 2.18.0
    functions {
    }
    data {
      int<lower=1> N;  // total number of observations
      vector[N] Y;  // response variable
      vector<lower=0>[N] weights;  // model weights
      int<lower=1> K;  // number of population-level effects
      matrix[N, K] X;  // population-level design matrix
      int prior_only;  // should the likelihood be ignored?
    }
    transformed data {
    }
    parameters {
      vector[K] b;  // population-level effects
      real<lower=0> sigma;  // dispersion parameter
    }
    transformed parameters {
      real lprior = 0;  // prior contributions to the log posterior
      lprior += student_t_lpdf(sigma | 3, 0, 137.9)
        - 1 * student_t_lccdf(0 | 3, 0, 137.9);
    }
    model {
      // likelihood including constants
      if (!prior_only) {
        // initialize linear predictor term
        vector[N] mu = rep_vector(0.0, N);
        mu += X * b;
        for (n in 1:N) {
          target += weights[n] * (normal_lpdf(Y[n] | mu[n], sigma));
        }
      }
      // priors including constants
      target += lprior;
    }
    generated quantities {
    }

\end{lstlisting}
    
  \item Plot the results: We then can plot the model results either for all parameters or for a specific subset. We compare the initial weighted posterior results (orange) with the adjusted weighted posterior (blue).
\begin{lstlisting}

    plot(mod.brms)
    plot(mod.brms, varnames = paste("b", 1:4, sep =""))

    \end{lstlisting}

    \begin{center}\includegraphics[width=.45\linewidth]{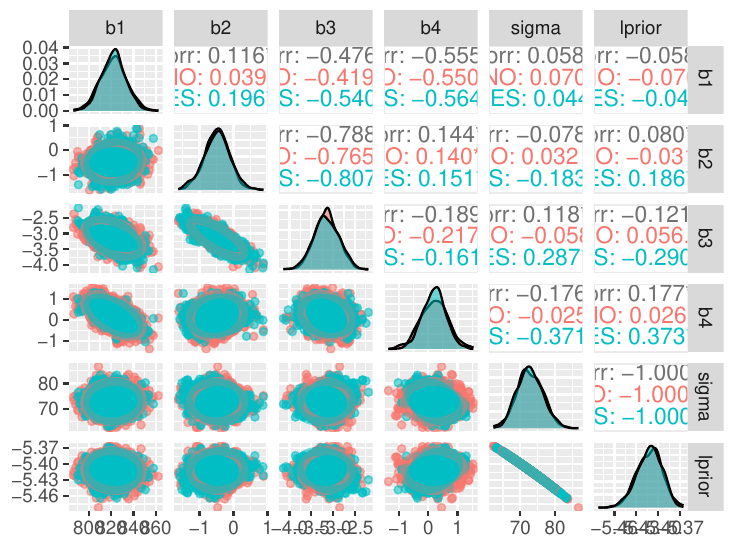} \includegraphics[width=.45\linewidth]{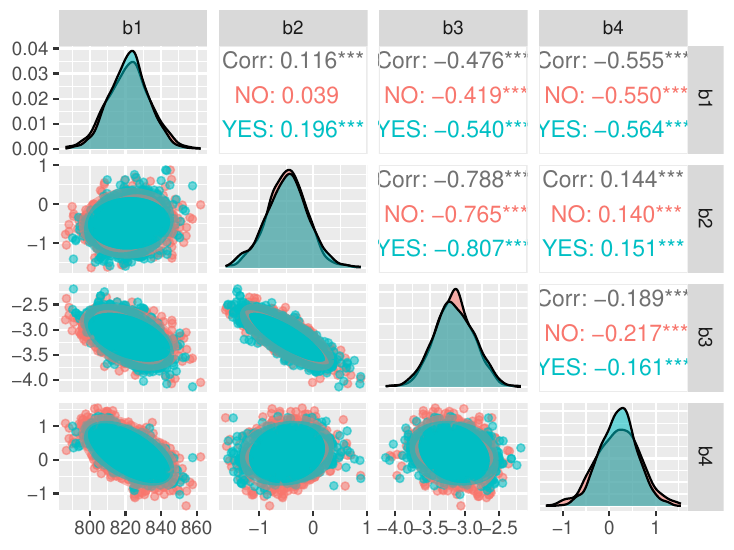} \end{center}
\end{enumerate}
In this simple regression example, we see little difference between the unadjusted and adjusted (pseudo-) posterior results. This study has both stratification (which tends to make results more efficient - narrower distributions) and unequal weighting of schools (which can make results less efficient - wider distributions - if weights are only weakly related to outcomes). In this example, these two factors have more or less canceled each other out, but differences are present across parameters. For example for English Language Learners (\texttt{ell}), the \texttt{b2} coefficient has a slightly wider distribution after adjustment (reflecting a design effect $> 1$) while for \texttt{mobility}, the \texttt{b4} coefficient has a slightly narrower distribution after adjustment (reflecting a design effect < 1).

\hypertarget{example-2}{%
\subsection{Example 2: multinomial dependent variable
}\label{example-2}}

In this example, we use \texttt{apiclus1} of the \texttt{api} data sets from the {\bf survey} R package. The \texttt{apiclus1} is a cluster sample of 183 school districts of California. In this example, we highlight the use of \texttt{cs\_sampling} function of {\bf csSampling} for use with a custom Stan model. Suppose we would like to estimate the proportion of school types \texttt{stype} (Elementary, Middle, High school) based on the clustered sample of school districts. This corresponds to the following population model:

\[
\begin{array}{rl}
    y_i|\theta &  \sim \ Multi(\theta) \\
    \theta & \sim \ D(\alpha) \\
    \sum_k \theta_k & = \  1 \\
\end{array}
\]
where each $y_i$ is an indicator variable across $k$ categories each with probability $\theta_k$. Here we choose a Dirichlet prior distribution for $\theta$ which lies on a simplex. 

For illustration, we choose to use the following equivalent formulation which allows us to work with an unrestricted vector of parameters $\lambda$ rather than a simplex $\theta$. This is the model we choose to include in Stan below.
\[
\begin{array}{rl}
    y_i|\theta &  \sim \ Multi(\theta) \\
    \theta_k & = \ \lambda_k / \sum_j \lambda_j \\
    \lambda_k & \sim \ G(\alpha, 1)
\end{array}
\]
where $G$ is the gamma distribution with scale $\alpha$ and shape $1$.
This parameterization is conceptually closer to what Stan does internally (using $\log (\lambda_k)$) for its unconstrained parameterization.

We estimate the survey-weighted Dirichlet-multinomial model following the steps below: 

\begin{enumerate}
  \item Define data and parameters of the complex sample: After normalizing the survey weight \texttt{pw} and defining \texttt{dclus1}, the complex sample parameters using \texttt{apiclus1}, we use \texttt{as.svrepdesign} to create a replicate-weights survey design object from \texttt{dclus1}. We use the default setting in \texttt{as.svrepdesign} which creates a set of jackknife replication weights to illustrate that alternative replication methods can be used.

\begin{lstlisting}

    data(api)
    apiclus1$wt <- apiclus1$pw/mean(apiclus1$pw)
    dclus1<-svydesign(id=~dnum, 
    +                    weights=~wt, 
    +                    data=apiclus1, 
    +                    fpc=~fpc)
    rclus1<-as.svrepdesign(dclus1)
    
    \end{lstlisting}
Using the \texttt{dclust1} design with other functions in the {\bf survey} package will estimate variances via the Taylor linearization approach. Using the \texttt{rclust1} design will lead to estimates of variance via replication methods.

   \item Construct a Stan model: use \texttt{stan\_model} to construct a Stan model based on model code defined in \texttt{proportion\_estimate} of {\bf csSampling}  
   
\begin{lstlisting}

    mod_dm <- stan_model(model_code = load_wt_multi_model())
    
    \end{lstlisting}
   
   The Stan model was not available to be generated easily with {\bf brms} because the unweighted version is conjugate and thus MCMC methods are not needed. For convenience, we bundle it with {\bf csSampling} as the function \texttt{load\_wt\_multi\_model()}:
   
\begin{lstlisting}

functions{
real wt_multinomial_lpmf(int[,] y, vector lambda, vector weights, 
                        int n, int K)
  {
    vector[K] theta;
    real check_term;
	int tmpy[K];
    theta = lambda / sum(lambda);
    check_term  = 0.0;
    for( i in 1:n )
    {
	tmpy = y[i,:];
	check_term    = check_term + 
	                weights[i] *  multinomial_lpmf(tmpy | theta);
    }
    return check_term;
  }
} /* end function{} block */

data {
	int<lower=1> n;
	int<lower=0> K;
	int<lower=0, upper = 1> y[n,K];
	vector<lower=0>[n] weights;
	vector<lower=0>[K] alpha;
}

parameters {
  vector<lower=0>[K] lambda;
}
transformed parameters{
  simplex[K] theta = lambda / sum(lambda);
  vector[K] loglam = log(lambda);
}

model {
	//theta ~ dirichlet(alpha);
	lambda    ~ gamma(alpha, 1 );
	target += wt_multinomial_lpmf(y | lambda, weights, n, K);
}
    
    \end{lstlisting}

   \item Prepare data for Stan modelling: define inputs for the Stan model, such as dependent variable and weights. We note that we use a uniform prior on the proportions ($\alpha = 1$).

\begin{lstlisting}

    y <- as.factor(rclus1$variables$stype)
    yM <- model.matrix(~y -1)
    n <- dim(yM)[1]
    K <- dim(yM)[2]
    alpha<-rep(1,K)
    weights <- rclus1$pweights
    data_stan<-list("y"=yM,
    +                 "alpha"=alpha,
    +                 "K"=K, 
    +                 "n" = n, 
    +                 "weights" = weights)
    ctrl_stan<-list("chains"=1,"iter"=2000,"warmup"=1000,"thin"=1)
    
    \end{lstlisting}

   \item Run the Stan model: use \texttt{cs\_sampling} of {\bf csSampling} to run a Stan model based on inputs defined in the previous step.  
   
\begin{lstlisting}

    mod1 <- cs_sampling(svydes = rclus1, 
    +                     mod_stan = mod_dm, 
    +                     data_stan = data_stan, 
    +                     ctrl_stan = ctrl_stan, 
    +                     rep_design = TRUE)
    
    \end{lstlisting}
      
    \item Plot the results: We then can plot the model results as below:
\begin{lstlisting}

    plot(mod1)
    plot(mod1, varnames = paste("theta", 1:3, sep =""))
    
    \end{lstlisting}

        \begin{center}\includegraphics[width=.45\linewidth]{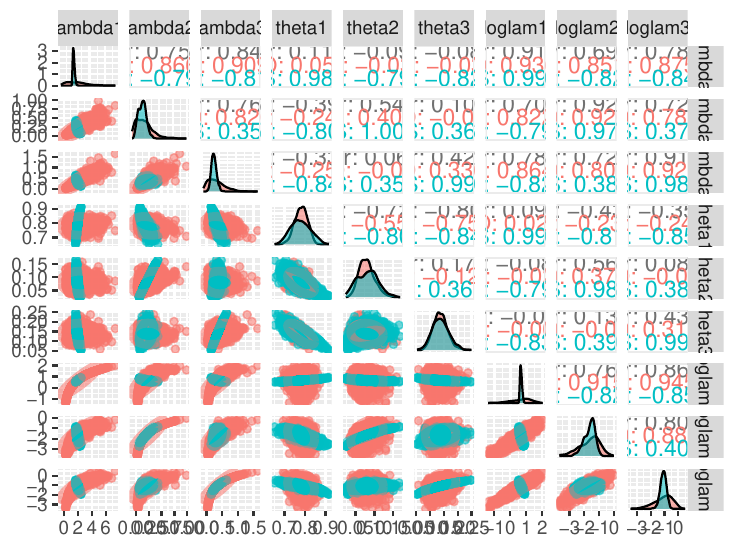} \includegraphics[width=.45\linewidth]{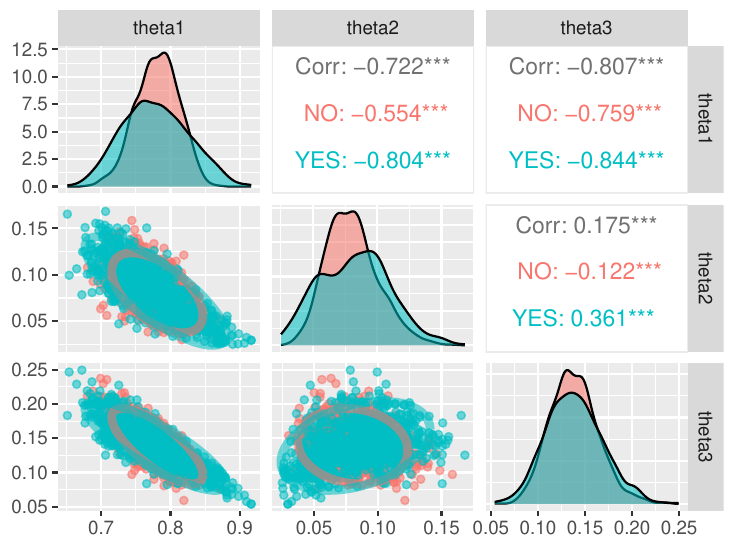} \end{center}
    We see that across all three school types, the adjusted posterior is more disperse (wider distribution - design effect $> 1$). This is more prominent for the estimate of elementary schools (\texttt{theta1}) and less so for high schools (\texttt{theta3}).

     \item Compare summary statistics: We compare between results from the Stan model and weighted percentage distribution using \texttt{svymean} of {\bf survey}.

    \begin{itemize}
    \item Results from Stan
    
\begin{lstlisting}

    cbind(colMeans(mod1$adjusted_parms[,4:6]), 
    +       sqrt(diag(cov(mod1$adjusted_parms[,4:6]))))
    
    \begin{CodeOutput}
                 [,1]       [,2]
    theta1 0.77906725 0.04471070
    theta2 0.08205868 0.02578196
    theta3 0.13887408 0.02937793
    \end{CodeOutput}
    \end{lstlisting}

    \item Results from \texttt{svymean} using replication
\begin{lstlisting}

    svymean(~stype, rclus1)
    
    \begin{CodeOutput}
               mean     SE
    stypeE 0.786885 0.0514
    stypeH 0.076503 0.0278
    stypeM 0.136612 0.0332
    \end{CodeOutput}
    \end{lstlisting}
    
    \item Results from \texttt{svymean} using Taylor linearization
\begin{lstlisting}

    svymean(~stype, dclus1)
    
    \begin{CodeOutput}
               mean     SE
    stypeE 0.786885 0.0463
    stypeH 0.076503 0.0268
    stypeM 0.136612 0.0296
    \end{CodeOutput}
    \end{lstlisting}

    \end{itemize}
    We see that the adjusted survey-weighted posterior estimates line up well with the survey-weighted mean estimates, matching closely when rounding to the nearest percentage. We also note that the uniform prior will pull each estimate closer to $1/3$, which we see in the results: \texttt{theta1} is shrunk down slightly while \texttt{theta2} and \texttt{theta3} are slightly increased. We also note that the estimated standard errors are closer to the Taylor linearization based variance estimates (using design \texttt{dclust1}) than to the purely replication based ones (using design \texttt{rclust1}).
\end{enumerate}

\hypertarget{example-3}{%
\subsection{Example 3: Bernoulli-distributed dependent variable
}\label{example-3}}

In this example, we use the {\bf brms} helper functions directly and \texttt{cs\_sampling} instead of the \texttt{cs\_sampling\_brms} wrapper as an illustration. We use the \texttt{dat14} data set from {\bf csSampling}. The \texttt{dat14} is a subset from the National Survey on Drug Use and Health (NSDUH). Check \url{https://nsduhweb.rti.org/respweb/homepage.cfm/} for more details about NSDUH. In this example, we run a Stan model on adults 18+ years old (\texttt{CATAG} $> 1$). We model \texttt{CIGMON}, that is 1 if respondent smoked cigarettes in the past month and 0 otherwise, and use \texttt{AMDEY2\_U} as a predictor variable (\texttt{AMDEY2\_U}: respondent had a major depressive episode during the past year). Our population model is then a simple logistic regression:

\[
\begin{array}{rl}
    y_i|p_i &  \sim \ Ber(p_i) \\
    \log\left(\frac{p_i}{1-p_i}\right)  & = \ \mu_i \\
    \mu_i & = \ X_i \beta \\
    \beta & \sim \ \propto 1
\end{array}
\]
where we use the logit link function for the linear predictor $\mu_i$ and a flat (improper) prior for $\beta$.

In order to estimate the survey-weighted model, we run the following the steps: 

\begin{enumerate}
  \item Define the data and the parameters of complex sample: We first normalize the survey weight \texttt{ANALWT\_C} then define \texttt{svy14}, the sample design object, using the cluster \texttt{VEREP} and strata \texttt{VESTR} variables. By default, \texttt{cs\_sampling} will convert this into a half-sample bootstrap replicate design.  

\begin{lstlisting}

    dat14 <- csSampling::dat14
    dat14 <- dat14[as.numeric(dat14$CATAG6) > 1,]
    dat14$WTS <- dat14$ANALWT_C/mean(dat14$ANALWT_C))
    svy14 <- svydesign(ids = ~VEREP, 
    +                    strata = ~VESTR, 
    +                    weights = ~WTS, 
    +                    data = dat14, 
    +                    nest = TRUE)
    
    \end{lstlisting}

   \item Construct a Stan model: We use \texttt{make\_stancode} of {\bf brms} to generate Stan code for our models.
   
\begin{lstlisting}

    model_formula <- formula("CIGMON|weights(WTS) ~ AMDEY2_U")
    stancode <- make_stancode(brmsformula(model_formula,center = FALSE), 
    +               data = dat14, 
    +               family = bernoulli(), 
    +               save_model = "brms_wt_log.stan")
    mod_brms  <- stan_model('brms_wt_log.stan')
    
    \end{lstlisting}
    
The generated Stan model for weighted logistic regression is then

\begin{lstlisting}

// generated with brms 2.18.0
functions {
}
data {
  int<lower=1> N;  // total number of observations
  int Y[N];  // response variable
  vector<lower=0>[N] weights;  // model weights
  int<lower=1> K;  // number of population-level effects
  matrix[N, K] X;  // population-level design matrix
  int prior_only;  // should the likelihood be ignored?
}
transformed data {
}
parameters {
  vector[K] b;  // population-level effects
}
transformed parameters {
  real lprior = 0;  // prior contributions to the log posterior
}
model {
  // likelihood including constants
  if (!prior_only) {
    // initialize linear predictor term
    vector[N] mu = rep_vector(0.0, N);
    mu += X * b;
    for (n in 1:N) {
      target += weights[n] * (bernoulli_logit_lpmf(Y[n] | mu[n]));
    }
  }
  // priors including constants
  target += lprior;
}
generated quantities {
}
    
    \end{lstlisting}
   
   \item Prepare data for Stan modelling: We use \texttt{make\_standata} of {\bf brms} to create a list of data inputs for the Stan model.
   
\begin{lstlisting}

    data_brms <- make_standata(brmsformula(model_formula,center = FALSE), 
    +                           data = dat14, 
    +                           family = bernoulli())
    
    \end{lstlisting}

   \item Run the Stan model and perform the adjustment using the default settings:  
   
\begin{lstlisting}

    mod.brms <- cs_sampling(svydes = svy14, 
    +                           mod_stan = mod_brms, 
    +                           data_stan = data_brms)
    
    \end{lstlisting}
      
    \item Plot the results:
\begin{lstlisting}

    plot(mod.brms, varnames = paste("beta", 1:2, sep =""))
    
    \end{lstlisting}
    
    \begin{center}\includegraphics[width=.45\linewidth]{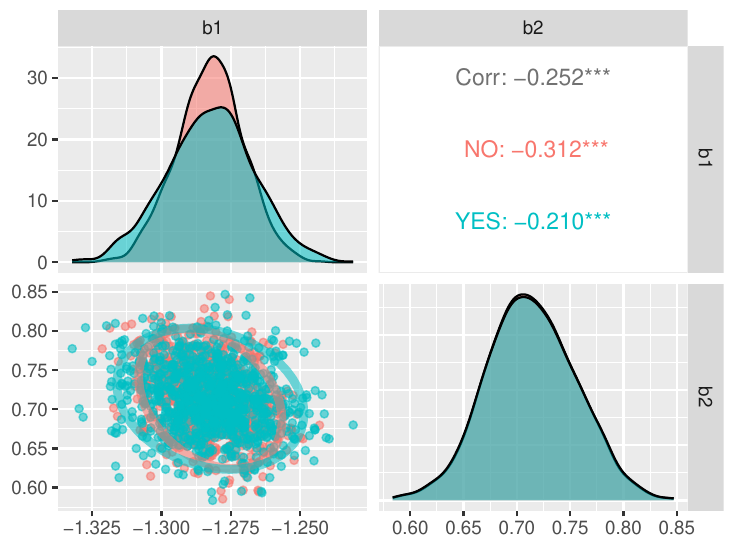} \end{center}
    
    We see that the intercept term \texttt{b1} (corresponding to no past year major depressive episode (MDE) has an adjusted posterior that is more disperse (design effect $> 1$). In contrast the regression `slope' \texttt{b2} corresponding to the difference between presence and absence of past year MDE is essentially unchanged (design effect $= 1$).

\end{enumerate}   

\bibliography{oct_2022}
\bibliographystyle{chicago}

\end{document}